\PassOptionsToPackage{usenames,dvipsnames,svgnames,table}{xcolor}
\RequirePackage{fix-cm}
\documentclass[twocolumn]{svjour3}          % twocolumn
\smartqed  % flush right qed marks, e.g. at end of proof

\usepackage[T1]{fontenc}
\usepackage[utf8]{inputenc}
\usepackage[left]{lineno}
\usepackage[draft]{hyperref} 
\usepackage{amsmath}
\usepackage{amssymb}
\usepackage{graphicx}
\graphicspath{{images/}}
\usepackage{subcaption}
\usepackage{booktabs}
\usepackage{transparent}
\pdfoutput=1 

\usepackage[]{todonotes}

\usepackage{multicol}
\usepackage{units}
\usepackage[super]{nth}
\usepackage{hepnames}
\newcommand{\PUpsilonFiveS}{\PUpsilon (5 \mathrm{S})}

\usepackage[numbers,sort&compress]{natbib}

\usepackage{authblk}
\usepackage{xcolor}
\usepackage{tikz}
\usetikzlibrary{calc}
\usetikzlibrary{chains}
\usetikzlibrary{positioning,arrows}
\usetikzlibrary{decorations.pathmorphing}
\usetikzlibrary{decorations.markings}
\usetikzlibrary{fadings,positioning,through}

\newcommand*{\belowrulesepcolor}{%
  \noalign{%
    \kern-\belowrulesep
    \begingroup
      \color{HeadRowColor}%
      \hrule height\belowrulesep
    \endgroup
  }%
}
\newcommand*{\aboverulesepcolor}{%
  \noalign{%
    \begingroup
      \color{HeadRowColor}%
      \hrule height\aboverulesep
    \endgroup
    \kern-\aboverulesep
  }%
}

\usepackage{listings}
\definecolor{Tblue}{HTML}{3465A4}	% tango sky blue 2
\definecolor{Tbluedark}{HTML}{204A87}	% tango sky blue 3
\definecolor{Tbluelight}{HTML}{729FCF}	% tango sky blue 1
\definecolor{Tbluelighter}{HTML}{8CC4FF}	% own definition of lighter blue
\definecolor{Tbrown}{HTML}{C17D11}	% tango chocolate 2
\definecolor{Tbrowndark}{HTML}{8F5902}	% tango chocolate 3
\definecolor{Tbrownlight}{HTML}{E9B96E}	% tango chocolate 1
\definecolor{Tgray}{HTML}{888A85}	% tango aluminium 4
\definecolor{Tgraydark}{HTML}{555753}	% tango aluminium 5
\definecolor{Tgraydarker}{HTML}{2E3436}	% tango aluminium 5
\definecolor{Tgraylight}{HTML}{BABDB6}	% tango aluminium 3
\definecolor{Tgraylight2}{HTML}{E4E6E2}	% Sehr hell (für Tabellenköpfe)
\definecolor{Tgraylight3}{HTML}{F0F2EE}	% Sehr hell (für Quelltexte)
\definecolor{Tgreen}{HTML}{73D216}	% tango chameleon 2	
\definecolor{Tgreendark}{HTML}{4E9A06}	% tango chameleon 3
\definecolor{Tgreenlight}{HTML}{8AE234}	% tango chameleon 1
\definecolor{Tred}{HTML}{CC0000}	% tango scarlet red 2
\definecolor{Treddark}{HTML}{A40000}	% tango scarlet red 3
\definecolor{Tredlight}{HTML}{EF2929}	% tango scarlet red 1	
\definecolor{Tlilac}{HTML}{75507B}	% tango plum 2
\definecolor{Tlilacdark}{HTML}{5C3566}	% tango plum 3
\definecolor{Tlilaclight}{HTML}{AD7FA8}	% tango plum 1
\definecolor{Tyellow}{HTML}{EDD400}	% tango butter 2
\definecolor{Tyellowdark}{HTML}{C4A000}	% tango butter 1
\definecolor{Tyellowlight}{HTML}{FCE94F}% tango butter 3
\definecolor{Torange}{HTML}{F57900}	% tango orange 2
\definecolor{Torangedark}{HTML}{CE5C00}	% tango orange 1
\definecolor{Torangelight}{HTML}{FCAF3E}% tango orange 3

\definecolor{Tgrayforkitprinter}{HTML}{EEEEEE}	% Sehr hell (für Tabellenköpfe)
\colorlet{HeadRowColor}{Tgrayforkitprinter}
\colorlet{CrossFeedColor}{Tblue}

\begin{document}

\title{B2BII}
\subtitle{Data conversion from Belle to Belle~II} 

\author{Moritz Gelb$^1$ \and Thomas Keck$^1$ \and Markus Prim$^1$ \and Hulya Atmacan$^2$ \and Jochen Gemmler$^1$ \and Ryosuke Itoh$^3$ \and Bastian Kronenbitter$^{1\star}$ \and Thomas Kuhr$^4$ \and Matic Lubej$^5$ \and Felix Metzner$^1$  \and Chanseok Park$^6$  \and Seokhee Park$^6$ \and Christian Pulvermacher$^{1\star}$ \and Martin Ritter$^4$ \and Anze Zupanc$^{5\star}$ }

%Seokhee Park (S-H.~Park) Yonsei University
%Chanseok Park (C.-S. Park) Yonsei University
%Hulya Atmacan University of South Carolina
%Thomas Kuhr 

\institute{T. Keck \at
			Karlsruhe Institute of Technology\\
			Institute of Experimental Particle Physics \\
		    Wolfgang-Gaede-Str. 1 \\
		    76131 Karlsruhe \\
            \email{thomas.keck2@kit.edu} 
            \and
            $^1$ Karlsruhe Institute of Technology \\
            $^2$ University of South Carolina \\
            $^3$ High Energy Accelerator Research Organization (KEK) \\
            $^4$ Ludwig Maximilians University Munich \\
            $^5$ University of Ljubljana \\
            $^6$ Yonsei University \\
            $^\star$ Research was performed while the author was affiliated with the corresponding institute
}

\date{Received: date / Accepted: date}

\authorrunning{B2BII}

\maketitle 

\begin{abstract}
We describe the conversion of simulated and recorded data by the Belle experiment to the Belle~II format with the software package \texttt{b2bii}. It is part of the Belle~II Analysis Software Framework. This allows the validation of the analysis software and the improvement of analyses based on the recorded Belle dataset using newly developed analysis tools.

\keywords{b2bii, belle, belle2, hep, basf2, data preservation}
\end{abstract}

\section{Introduction}
The Belle experiment recorded a dataset of approximately $\unit[1]{ab^{-1}}$ during its runtime; mainly at the $\PUpsilonFourS$ resonance. The physics program was very successful with milestones such as the measurement of mixing-induced CPV in $B^0 \to J/\Psi K_S^0$ decays leading to the Noble Prize for Kobayashi and Maskawa in 2008 \citep{PhysRevD.66.032007}, the precise measurement of the CKM matrix elements \citep{Amhis:2016xyh}, and the discovery of tetra quarks \citep{PhysRevLett.100.142001}. \\
Its successor, the Belle~II experiment, will soon start to record the first collisions. To allow for the envisaged 40-times higher peak luminosity, the collider and detector were upgraded. In addition, the Belle~II Analysis Software Framework (\texttt{BASF2}) \citep{basf2} was developed from scratch. An thorough validation of the software is necessary to ensure the integrity of upcoming analyses.\\
In this article we describe the software package (\texttt{b2bii}) based on \citep{ThomasPhd}, which converts simulated and recorded Belle events into the Belle~II format. 

\subsection{The Belle \& Belle~II Detector}
The design of the Belle~II detector resembles it predecessor. Each individual sub-detector is upgraded with a modern version of itself. For a detailed description of the Belle and Belle~II detectors see reference~\cite{Abashian2002117} and~\cite{B2TDR}, respectively. 

Going outwards from the interaction point (IP) the Belle detector consisted of a four layer silicon strip detector (SVD), a central drift chamber (CDC), an Aerogel Cherenkov counter (ACC), a time-of-flight (TOF) detector system, an electromagnetic calorimeter (ECL), a superconducting solenoid which provided a homogeneous magnetic field of $\unit[1.5]{T}$, and a return yoke, which was instrumented with glass-electrode resistive plate counters for $K_L$ and muon detection (KLM).

Going outwards from the IP the Belle~II detector consists of a two layer pixel detector (PXD), a four layer silicon strip detector (SVD), a central drift chamber (CDC), a proximity-focusing Aerogel ring-imaging Cherenkov detector (ARICH), a time-of-propagation counter (TOP), an electromagnetic calorimeter (ECL), a superconducting solenoid which provides a homogeneous magnetic field of $\unit[1.5]{T}$, and a return yoke, which is instrumented with glass-electrode resistive plate counters in the barrel region and scintillator strips in the end-caps for $K_L$ and muon detection (KLM).

\subsection{Recorded Belle Data}
Most of the Belle data was recorded at the center-of-mass energy of the $\PUpsilonFourS$ resonance. In addition, data was also recorded at the $\PUpsilonOneS$, $\PUpsilonTwoS$, $\PUpsilonThreeS$ and $\PUpsilonFiveS$ resonances. Moreover, off-resonance data, mostly used to study non-resonant background processes, was recorded.

The raw data coming from the detector was calibrated, reconstructed and stored on tape using \texttt{PANTHER}-based data summary tape (DST) files. \texttt{PANTHER} is a custom serialization format \citep{KATAYAMA199822}. After each experiment the calibration constants were recomputed by detector experts or computed directly from data, and stored in the Belle Condition Database, based on \texttt{PostgreSQL}.
Finally, the data of the completed experiment was reprocessed and stored in a compact form called mDST files, a reduced and compressed form of the data summary tape files.
The reconstruction and the processing of the mDST files is handled by the Belle Analysis Framework (\texttt{BASF}) \cite{basf}.
Different types of events were simulated using the \texttt{EvtGen} \citep{evtgen} and \texttt{GEANT3} \citep{Geant3} packages, and reconstructed with the same software as was used for the detector data.

\subsection{Anticipated Belle~II Data}
%The original detector is currently upgraded to match the higher instantaneous luminosity of SuperKEKB.
% anticip. dataset
By 2025, Belle~II will record $\unit[50]{ab^{-1}}$ of data, which corresponds to 50 times the integrated luminosity of Belle. The same software framework is used in online data taking and offline reconstruction, Monte Carlo production, and physics analysis. After time-dependent calibration parameters are determined, the raw data is reconstructed and stored at the KEK computing center\footnote{Other computing centers will store additional copies of the raw data.}. The time-dependent calibration parameters are stored in the Belle~II Condition Database \citep{RitterConditionDB} \citep{belle2_db}. Monte Carlo production and reconstruction will be distributed to data centers around the world. The reconstructed information is stored in \texttt{ROOT}-based \citep{Brun199781} mDST files. 

\subsection{Data Processing Levels}
\label{sec:b2bii:conversion_dataset}
In the above discussion of the recorded Belle and anticipated Belle~II dataset, four levels of data processing can be distinguished:
\begin{enumerate}
\item \textbf{online reconstruction} -- the read-out of the detector and the trigger system, producing the \textbf{raw-data} (DST files);
\item \textbf{offline reconstruction} -- cluster reconstruction, track finding and fitting, producing the \textbf{mDST}\\ \textbf{data};
\item \textbf{mDST analysis} -- creation of final state particle hypotheses, reconstruction of intermediate particle candidates
and vertex fitting, producing \textbf{flat n-tuples};
\item and \textbf{n-tuple analysis} -- fit to theoretical predictions in order to extract interesting observables, producing \textbf{scientific papers}.
\end{enumerate}

Converting the raw-data is in principle possible, but the differences between the
Belle and Belle~II detector render this a difficult and ill-defined task.
While this would allow for the validation of the Belle~II reconstruction software (e.g. the track finding
and fitting algorithms) on Belle data, this would be only of limited use due to the differences between the Belle and the Belle~II detector,
the vastly different expected background, and the availability of events recorded by Belle~II from cosmic runs.

The Belle to Belle~II dataset conversion converts the Belle mDST data, which contains mostly detector independent objects like tracks and calorimeter clusters, into the new mDST format used by \texttt{BASF2}. This enables the validation of the Belle~II analysis software, and (re-)production of Belle measurements using the improved software.

By comparing the original Belle results, the results obtained from converted data in \texttt{BASF2}, and Belle~II sensitivity studies on Belle~II Monte Carlo, it is possible to assign improvements in the sensitivity and occurring issues to the analysis and reconstruction algorithms, separately.

The Belle experiment provides a large amount of Monte Carlo simulated events, which can be processed using \texttt{b2bii}.
However, the production of additional Monte Carlo simulated Belle events still requires \texttt{BASF} and is not part of \texttt{b2bii}.

\section{Method}
\label{subsec:b2bii:overview}
\begin{figure*}
\centering
\begin{tikzpicture}[%
      oldbox/.style={
          rectangle,
          draw=black,
          thick, font=\normalsize,
          fill=Tbluelighter!50,
          postaction={path fading=north, fading angle=-45, fill=Tbluelight!50},
          align=center,
          minimum height=3.5em,
          minimum width=6em
        },
    newbox/.style={
          rectangle,
          draw=black,
          thick, font=\normalsize,
          fill=Tbluelighter,
          postaction={path fading=north, fading angle=-45, fill=Tbluelight},
          align=center,
          minimum height=3.5em,
          minimum width=6em
        },
    modbox/.style={
          rectangle,
          draw=black,
          thick, font=\normalsize,
          fill=Tgraylight2,
          postaction={path fading=north, fading angle=-45, fill=Tgraylight},
          align=center,
          minimum height=3.5em,
          minimum width=6em
        },
    libbox/.style={
          rectangle,
          draw=black,
          thick, font=\normalsize,
          fill=Tgraylight2,
          postaction={path fading=north, fading angle=-45, fill=Tgraylight},
          align=center,
          minimum height=3.5em,
          minimum width=24em
        },
            arrow/.style={
               ->,
               >=latex',
               line width=1.5pt,
               Tgraydarker,},
            ]
\draw (1.5,3.2) node[oldbox] (Mdst) {Belle\\ mDST-files};
\draw (0,-1) node[oldbox] (DB) {Belle\\ database};

\draw (4.5,3.2) node[oldbox] (Panther) {Panther\\ tables};
\draw (3,-1) node[modbox] (Input) {B2BII\\ MdstInput};
\draw (7.5,3.2) node[oldbox] (FixedPanther) {,,fixed'' Panther\\ tables};
\draw (6,-1) node[modbox] (FixMdst) {B2BII\\ FixMdst};
\draw (9,-1) node[modbox] (Convert) {B2BII\\ ConvertMdst};

\draw (10.5,3.2) node[newbox] (Mdst2) {Belle~II\\ mDST-files};
\draw (12,-1) node[newbox] (DB2) {Belle~II beam\\ parameters};

\draw (12,1) node[newbox] (DB3) {Belle~II\\ database};
\draw[arrow,dashed] (DB2.north) -> (DB3.south);

\draw[arrow] (Mdst.south) -> (Input.north);
\draw[arrow] (Input.north) -> (Panther.south);
\draw[arrow] (Panther.south) -> (FixMdst.north);
\draw[arrow] (FixMdst.north) -> (FixedPanther.south);
\draw[arrow] (FixedPanther.south) -> (Convert.north);
\draw[arrow] (Convert.north) -> (Mdst2.south);
\draw[arrow] (DB.east) -> (Input.west);
\draw[arrow] (Input.east) -> (FixMdst.west);
\draw[arrow] (FixMdst.east) -> (Convert.west);
\draw[arrow] (Convert.east) -> (DB2.west);

\draw (4.8, 1.1) node[libbox] (BelleLegacy) {\texttt{belle\_legacy} library};

\end{tikzpicture}
\caption{Schematic view of the conversion process of Belle (\textcolor{Tbluelight}{light blue}) to Belle~II (\textcolor{Tbluedark}{blue}) mDST files using the \texttt{BASF2} modules (\textcolor{Tgraydark}{gray}) provided by the \texttt{b2bii} package
and the original Belle software provided by the \texttt{belle\_legacy} library (\textcolor{Tgraydark}{gray}).}
\label{fig:b2bii:B2BIIChain}
\end{figure*}
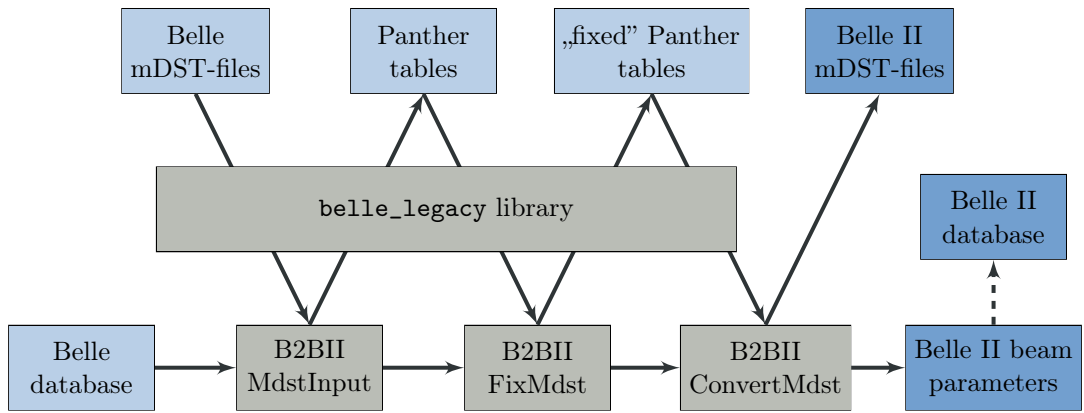

The software responsible for reading in the old \texttt{PANTHER} data-format and representing the data in memory was isolated, cleaned up and 
compiled into a new library named \texttt{belle\_legacy}. A new package was introduced in \texttt{BASF2} called \texttt{b2bii} (Belle to Belle~II). It contains three \texttt{BASF2} modules developed with the help of the \texttt{belle\_legacy} library.
The conversion process is visualized in \autoref{fig:b2bii:B2BIIChain}.

\begin{description}
\item [The \texttt{B2BIIMdstInput} module] opens the \texttt{PANTHER}-based Belle mDST files and reads the data event-by-event into the main memory. The data of the current event is represented in the memory by a series of \texttt{PANTHER} tables.
\item [The \texttt{B2BIIFixMdst} module] applies various calibration factors onto the \texttt{PANTHER} tables, for instance on the beam-energy, the momenta and error matrices of the fitted tracks, the energy deposition in the ECL, and the particle identification information of the CDC and TOF. It also performs standard cuts to ensure that the selection of the detector data and simulated events is identical to the one obtained with \texttt{BASF}. Finally, $\Ppizero$ candidates are reconstructed from the $\Pgamma$ particle objects and the corrected ECL clusters. An equivalent module named \texttt{FixMdst} was already used by \texttt{BASF}. 
\item [The \texttt{B2BIIConvertMdst} module] converts the information stored in the Belle \texttt{PANTHER} tables into the corresponding Belle~II \texttt{ROOT} objects.
The beam-energy and IP-profile is collected in the \texttt{BASF2}\\ \texttt{BeamParameters} object and stored in the condition Database of Belle~II.
\end{description}

\subsection{Data Formats}
\label{subsec:dataformats}
The Belle data format is based on a custom serialization format called \texttt{PANTHER}.
It consists of tables compressed by the \texttt{zlib} library. The table formats are defined by \texttt{ASCII} header files.
Each table consists of multiple rows, called entries. The index of each entry can be used to relate entries to other entries, for instance, to express a mother--daughter relationship between particles (a particle which decays inside the detector is called the mother
of its decay-products, which are the daughters).
\texttt{BASF} processes the data event-by-event, meaning the in-memory {representation} of the \texttt{PANTHER} tables contains only a single event at a given point.
After each event the tables are flushed and filled with the next event.

The Belle~II data format is based on \texttt{ROOT}. The \texttt{ROOT} framework takes care of serialization including potential migrations to ensure backward-compatibility. Conceptually we distinguish different types of \texttt{ROOT} objects.
\textbf{Array objects} are the equivalent to the tables used by \texttt{PANTHER}.
The entries of different array objects can be connected by adding so-called relations, which are stored in a separate array object.
The relations allow the expression of many-to-many connections between arbitrary entries of the array objects,
for instance the relation between a track and the associated clusters: this allows the analyst to easily access all clusters which are associated to a given track.
\textbf{Single objects} are used to store the remaining event-wise information, for instance the meta data of each event, or particle lists created by the analyst. A particle list is a list of \texttt{Particle} array entries used to organize the reconstruction of decay-chains in \texttt{BASF2}.
\texttt{BASF2} processes the data event-by-event, meaning the in-memory {representation} of most \texttt{ROOT} objects contains only a single event at a given point.
After each event the objects are filled with the next event.
Some \texttt{ROOT} objects are persistent in the sense that they are only stored and loaded once per file, for instance, the meta data of each file or statistics of the execution time and memory consumption used for profiling.

\subsection{Implementation Details}
\label{subsec:b2bii:implementation_details}
The detailed matching between \texttt{PANTHER} tables and corresponding \texttt{BASF2} data-objects is shown in \autoref{fig:b2bii:B2BIIConversion}.
In the following we describe the conversion process in detail for future reference.

\subsubsection{Event Information}
Event information like the beam energy and position of the IP are loaded from the Belle condition database and stored in \texttt{BeamParameters} objects that can be uploaded to the Belle~II condition database. The \texttt{BeamParameters} of the entire detector data was converted and uploaded. The \texttt{BeamParameters} of simulated events are only stored on the local machine.

The description of the magnetic field differs between Belle and Belle~II. The conversion uses a magnetic field map which is consistent with the track parametrization in Belle data.

\subsubsection{Monte Carlo}
The Monte Carlo information of Belle is stored in the so-called \texttt{Gen\_hepevt} table. It contains the four momenta of the generated particles
and the indices of the mother and all daughter particles. The table is converted into an array of \texttt{MCParticle} objects, which contains the same information.
Consequently, the fine-grained unified Monte Carlo matching algorithm of \texttt{BASF2} can be used, and problems contained in algorithms used by \texttt{BASF} are avoided \cite[sec.~4.3]{Pulvermacher}.

The \texttt{Gen\_hepevt} table includes special entries for a common mother of beam-background particles (PDG code 911) and for virtual photons (PDG code 0).
These entries are ignored during the conversion, because there are no corresponding concepts in Belle~II.
For instance, in \texttt{BASF2} beam-background is indicated by a motherless Monte Carlo particle.

The original Belle software does not provide Monte Carlo information for KLM clusters, following the approach of \cite[sec.~5.2]{KLongCalibration} true $\PKlong$ are matched to the closest reconstructed Monte Carlo $\PKlong$ within $\pm 15$ degrees in both $\theta$ and $\phi$.

Furthermore, unlike Belle~II simulated events, the Belle simulated events do not provide information on the differentiation between photons generated directly by the fundamental matrix-element calculated by the Monte Carlo generator \texttt{EvtGen} (hereinafter referred to as gamma) and photons generated afterwards for instance by \texttt{PHOTOS} \cite{Photos} or the simulation (hereinafter referred to as final state radiation) (see \cite[Appendix C]{EvtGenManuel}).
Often a reconstructed particle which misses final state radiation is considered as signal, whereas a reconstruction with a missing gamma is considered as wrong.
A simple heuristic is applied to distinguish the two cases:
Photons from a decay $M \rightarrow A B ... \Pgamma$ are flagged as final state radiation, and photons from a decay $M \rightarrow A \Pgamma$ are flagged as gammas.
In particular photons from $\Ppizero \rightarrow \Pgamma \Pgamma$ and $\PDstar \rightarrow \PD \gamma$ are considered gammas.
Other cases like $\PB \rightarrow \Pmu \Pnu \Pgamma$ are regarded by the heuristic as final state radiation and have to be treated by the analyst themself\footnote{In this case, photons from initial and final state radiation are physically indistinguishable, since the corresponding amplitudes interfere. Actually, there is no correct answer to the question of whether the photon is final state radiation or not. Hence, the behavior of the heuristic is not wrong, but probably
unexpected by the analyst, because the initial state radiation amplitude dominates in this decay.}.

The official Belle Monte Carlo campaigns produced ten times the real integrated luminosity in $\PB \APB$ events and six times that in continuum events,
however some inconsistencies were encountered during the development of the conversion software, which were fixed if possible:
The Monte Carlo campaign deleted the 8 left-most bits of the 32~bit long PDG codes during the Monte Carlo simulation\footnote{\texttt{BASF} already implemented a function for recovering the lost bits, but it was apparently not applied.}. During the conversion these corrupted PDG codes are restored by matching their lower 24~bit to known PDG codes.
In the official Belle Monte Carlo campaign from 2010 for $\PB \rightarrow \Pup \Plepton \Pnu$ and other rare $\PB$ decays, the mass of almost all MC particles is set to zero,
which can lead to wrong results if this quantity is used during the analysis. However, this information is redundant since the correct mass of the MC particles
can be calculated using either the PDG values or the MC four-momenta.

\subsubsection{Tracks}
The track reconstruction output of \texttt{BASF} is stored in the so-called \texttt{Mdst\_charged} and \texttt{Mdst\_trk} tables.
They contain the 5D track parametrization for up to five different final state particle hypotheses.\\ The track parametrization is transformed and stored into \texttt{Track} and associated \texttt{TrackFitResult} array objects. The transformation is unique but non-trivial because the two experiments employ different 5D track parameterizations and conventions for the reference point of the track.

\subsubsection{ECL Clusters}
The output of the ECL cluster algorithm of Belle is stored in the so-called \texttt{Mdst\_ecl} and \texttt{Mdst\_ecl\_aux} tables.
They contain information about the energy, position and shape of the clusters.
The ECL information is converted and stored in the \texttt{ECLCluster} array object.
Information is mapped to the corresponding representation, e.g. the energy and position of the clusters with the corresponding covariance matrix and shower variables, such as the $E9E25$ ratio (which is stored in the field for the $E9E21$ ratio as this is the one now used in Belle~II). Advanced shower variables like Zernike moments were not available for Belle and are therefore not set.
In addition two \texttt{ParticleList} objects are created containing the $\Pgamma$ and $\Ppizero$ candidates, which were created by \texttt{B2BIIFixMdst} earlier. The lists are named \texttt{gamma:mdst} and \texttt{pi0:mdst}, respectively. The \texttt{ParticleList} objects provide a fast and easy access to the possible $\Pgamma$ and $\Ppizero$ candidates, used by the analyst during their analysis.

\subsubsection{KLM Clusters}
The output of the KLM cluster algorithm of Belle is stored in the so-called \texttt{Mdst\_klm\_cluster} and\\ \texttt{Mdst\_klong} tables.
The KLM information is converted and stored in the \texttt{KLMCluster} array object. In addition a \texttt{ParticleList} is filled containing $\PKlong$ candidates. The list is named \texttt{K\_L0:mdst}.

\subsubsection{V0 Objects}
A V0 object is a pair of tracks with a common vertex usually outside of the beam pipe. Such a signature indicates the decay of a particle with
a relatively long lifetime like a $\PKshort$. 
The output of the \texttt{V0 Finder} of Belle is stored in the so-called \texttt{Mdst\_vee\_daughters} and \texttt{Mdst\_vee} tables.
Additional information is created on-the-fly by the \texttt{nisKsFinder}, which provides quality information.
The V0 information  is directly transformed into \texttt{ParticleList} objects containing candidates for $\Pgamma$, $\PKshort$ and $\PLambda$.\\ The lists are named \texttt{gamma:v0mdst}, \texttt{K\_S0:mdst} and \texttt{Lambda0:mdst}, respectively.\\
The additional quality information is stored in the\\ \texttt{ExtraInfo} field of the \texttt{Particle} array entries
under the keys \texttt{goodKs}, \texttt{ksnbVLike}, \texttt{ksnbNoLam} and \texttt{ksnbStandard}.

\subsubsection{PID Information}
The PID information provided by the different Belle sub-detectors is stored in the so-called \texttt{kid\_acc},\\ \texttt{Mdst\_tof}. \texttt{kid\_cdc}, \texttt{eid} and \texttt{Mdst\_klm\_mu\_ex} tables. It is mapped to similar Belle~II sub-detectors, so that the physical meaning of the information is partially preserved. In particular the Belle time-of-flight (TOF) and Aerogel Cherenkov counter (ACC) detectors are mapped to the Belle~II time-of-propagation (TOP) and Aerogel ring imaging Cherenkov (ARICH) detectors, respectively. The converted information is stored in the \texttt{PIDLikelihood} array object.

\subsubsection{Relations}
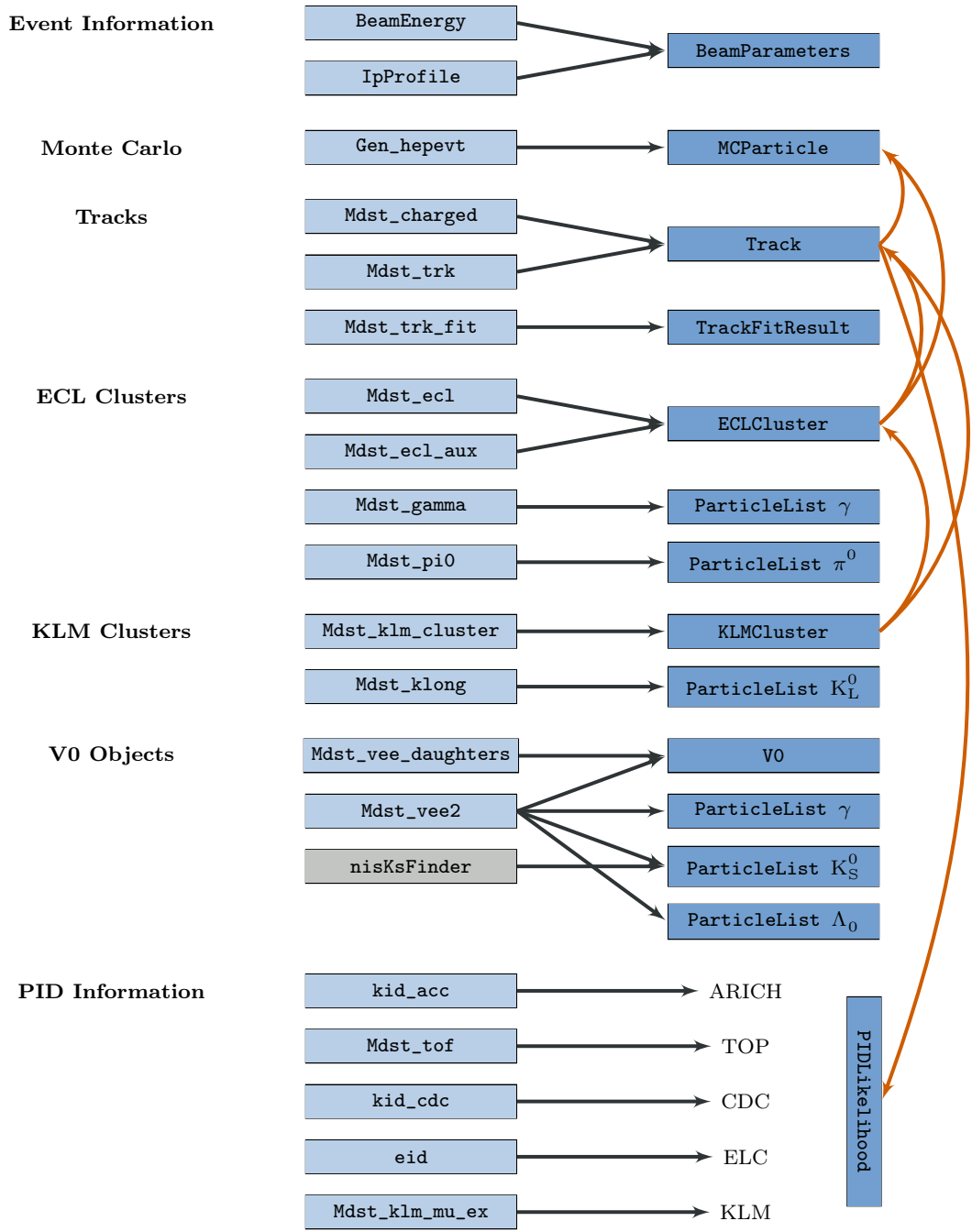
\begin{figure*}
\centering
\begin{tikzpicture}[%
b1box/.style={
      rectangle,
      draw=Black,
      thick,
          fill=Tbluelighter!50,
          postaction={path fading=north, fading angle=-45, fill=Tbluelight!50},
      align=center,
      font=\footnotesize\ttfamily,
      minimum height=1.5em,
      minimum width=9.5em
    },  
    ksbox/.style={
          rectangle,
          draw=Black,
          thick,
              fill=Tgray!50,
              postaction={path fading=north, fading angle=-45, fill=Tgray!50},
          align=center,
          font=\footnotesize\ttfamily,
          minimum height=1.5em,
          minimum width=9.5em
        },  
b2box/.style={
      rectangle,
      draw=Black,
      thick,
          fill=Tbluelighter,
          postaction={path fading=north, fading angle=-45, fill=Tbluelight},
      align=center,
      font=\footnotesize\ttfamily,
      minimum height=1.5em,
      minimum width=9.5em
    },  
arrow/.style={
   ->,
   >=latex',
   shorten >=1pt,
   ->,
   line width=1.5pt,
   Tgraydarker,},
relation/.style={
   ->,
   >=latex',
   shorten >=1pt,
   ->,
   line width=1.5pt,
   Torangedark,},
]

\draw (-4.3,2) node () {\textbf{Event Information}};
\draw (0,2) node[b1box] (BeamEnergy) {BeamEnergy};
\draw (5.2,1.6) node[b2box] (BeamParameters) {BeamParameters};
\draw (0,1.2) node[b1box] (IpProfile) {IpProfile};
\draw[arrow] (BeamEnergy.east) -> (BeamParameters.west);
\draw[arrow] (IpProfile.east) -> (BeamParameters.west);

\draw (-4.3,0.2) node () {\textbf{Monte Carlo}};
\draw (0,0.2) node[b1box] (GenHepEvt) {Gen\_hepevt};
\draw (5.2,0.2) node[b2box] (MCParticle) {MCParticle};
\draw[arrow] (GenHepEvt.east) -> (MCParticle.west);

\draw (-4.3,-0.8) node () {\textbf{Tracks}};
\draw (0,-0.8) node[b1box] (MdstCharged) {Mdst\_charged};
\draw (0,-1.6) node[b1box] (MdstTrk) {Mdst\_trk};
\draw (0,-2.4) node[b1box] (MdstTrkFit) {Mdst\_trk\_fit};
\draw (5.2,-1.2) node[b2box] (Track) {Track};
\draw (5.2,-2.4) node[b2box] (TrackFitResult) {TrackFitResult};

\draw[arrow] (MdstCharged.east) -> (Track.west);
\draw[arrow] (MdstTrk.east) -> (Track.west);
\draw[arrow] (MdstTrkFit.east) -> (TrackFitResult.west);

\draw (-4.3,-3.4) node () {\textbf{ECL Clusters}};
\draw (0,-3.4) node[b1box] (MdstEcl) {Mdst\_ecl};
\draw (0,-4.2) node[b1box] (MdstEclAux) {Mdst\_ecl\_aux};
\draw (0,-5.0) node[b1box] (MdstGamma) {Mdst\_gamma};
\draw (0,-5.8) node[b1box] (MdstPi0) {Mdst\_pi0};
\draw (5.2,-3.8) node[b2box] (ECLCluster) {ECLCluster};
\draw (5.2,-5.0) node[b2box] (ParticleListGammaMdst) {ParticleList $\Pgamma$};
\draw (5.2,-5.8) node[b2box] (ParticleListPi0Mdst) {ParticleList $\Pgpz$};

\draw[arrow] (MdstEcl.east) -> (ECLCluster.west);
\draw[arrow] (MdstEclAux.east) -> (ECLCluster.west);
\draw[arrow] (MdstGamma.east) -> (ParticleListGammaMdst.west);
\draw[arrow] (MdstPi0.east) -> (ParticleListPi0Mdst.west);

\draw (-4.3,-6.8) node () {\textbf{KLM Clusters}};
\draw (0,-6.8) node[b1box] (MdstKlm) {Mdst\_klm\_cluster};
\draw (0,-7.6) node[b1box] (MdstKlong) {Mdst\_klong};
\draw (5.2,-6.8) node[b2box] (KLMCluster) {KLMCluster};
\draw (5.2,-7.6) node[b2box] (ParticleListKlongMdst) {ParticleList $\PKlong$};

\draw[arrow] (MdstKlm.east) -> (KLMCluster.west);
\draw[arrow] (MdstKlong.east) -> (ParticleListKlongMdst.west);

\draw (-4.3,-8.6) node () {\textbf{V0 Objects}};
\draw (0,-8.6) node[b1box] (MdstVeeDaughters) {Mdst\_vee\_daughters};
\draw (0,-9.4) node[b1box] (MdstVee2) {Mdst\_vee2};
\draw (0,-10.2) node[ksbox] (nisKsFinder) {nisKsFinder};
\draw (5.2,-8.6) node[b2box] (V0) {V0};
\draw (5.2,-9.4) node[b2box] (ParticleListGammaV0) {ParticleList $\Pgamma$};
\draw (5.2,-10.2) node[b2box] (ParticleListKS0Mdst) {ParticleList $\PKshort$};
\draw (5.2,-11) node[b2box] (ParticleListLambda0Mdst) {ParticleList $\PLambda_0$};

\draw[arrow] (MdstVeeDaughters.east) -> (V0.west);
\draw[arrow] (MdstVee2.east) -> (V0.west);
\draw[arrow] (MdstVee2.east) -> (ParticleListKS0Mdst.west);
\draw[arrow] (nisKsFinder.east) -> (ParticleListKS0Mdst.west);
\draw[arrow] (MdstVee2.east) -> (ParticleListLambda0Mdst.west);
\draw[arrow] (MdstVee2.east) -> (ParticleListGammaV0.west);

\draw (-4.3,-12.0) node () {\textbf{PID Information}};
\draw (0,-12.0) node[b1box] (KIDACC) {kid\_acc};
\draw (0,-12.8) node[b1box] (MdstTOF) {Mdst\_tof};
\draw (0,-13.6) node[b1box] (KIDCDC) {kid\_cdc};
\draw (0,-14.4) node[b1box] (EID) {eid};
\draw (0,-15.2) node[b1box] (MdstKLMMuEx) {Mdst\_klm\_mu\_ex};
\draw (6.5,-13.6) node[b2box,rotate=-90] (PID) {PIDLikelihood};

\draw (4.8, -12.0) node (ARICH) {\footnotesize ARICH};
\draw (4.8, -12.8) node (TOP) {\footnotesize TOP};
\draw (4.8, -13.6) node (CDC) {\footnotesize CDC};
\draw (4.8, -14.4) node (ECL) {\footnotesize ELC};
\draw (4.8, -15.2) node (KLM) {\footnotesize KLM};
\draw[arrow] (KIDACC.east) -> (ARICH.west);
\draw[arrow] (MdstTOF.east) -> (TOP.west);
\draw[arrow] (KIDCDC.east) -> (CDC.west);
\draw[arrow] (EID.east) -> (ECL.west);
\draw[arrow] (MdstKLMMuEx.east) -> (KLM.west);

\draw[relation] (Track.east) to [bend right=50] (MCParticle.east);
\draw[relation] (Track.east) to [bend left=20] (PID.north);
\draw[relation] (ECLCluster.east) to [bend right=50] (MCParticle.east);
\draw[relation] (ECLCluster.east) to [bend right=50] (Track.east);
\draw[relation] (KLMCluster.east) to [bend right=50] (Track.east);
\draw[relation] (KLMCluster.east) to [bend right=50] (ECLCluster.east);

\end{tikzpicture}
\caption{Matching of the Belle \texttt{PANTHER} Tables (\textcolor{Tblue!50}{light blue}) to the Belle~II \texttt{ROOT} objects (\textcolor{Tbluedark}{blue}) and relations (\textcolor{Torangedark}{orange}).}
\label{fig:b2bii:B2BIIConversion}
\end{figure*}

Finally, some of the created array entries are related to one another (see \autoref{fig:b2bii:B2BIIConversion}). Hence, \texttt{BASF2} relations are created: from the \texttt{ECLCluster} entries to the \texttt{MCParticle} and \texttt{Track} entries which are responsible for the creation of the cluster; similarly from the \texttt{KLMCluster} entries to the \texttt{ECLCluster} and \texttt{Track} entries; from the \texttt{Track} entries to the \texttt{MCParticle} entries that created it; and from the \texttt{Track} entries to the corresponding \texttt{PIDLikelihood} entries.
Additional relations are created between the \texttt{Particle} entries in the created \texttt{ParticleList} objects and the corresponding \texttt{MCParticle} and \texttt{PIDLikelihood} entries.
The links between \texttt{TrackFitResult} to \texttt{Track}; \texttt{Track} to \texttt{Particle}; and \texttt{ECLCluster} and \texttt{KLMCluster} to \texttt{Particle} are not represented by relations in \texttt{BASF2}.

\section{Discussion}
\label{sec:b2bii:validation}
In order to ensure the correctness of the conversion, a study was performed with $200\,000$ recorded data and $600\,000$ simulated Belle events at the center of mass energy of the $\PUpsilonFourS$ resonance.

The events were processed with the old \texttt{BASF} framework and more than $360$ quantities; for instance kinematic quantities like four-momenta, Monte Carlo information, PID information and beam-parameters were extracted from the \texttt{PANTHER} tables shown in \autoref{fig:b2bii:B2BIIConversion}. The complete list of extracted quantities can be found in \citep[app.~A]{ThomasPhd}.
Afterwards the events were processed a second time with the new \texttt{BASF2} software using the \texttt{b2bii} conversion and the same quantities were extracted.

\subsection{Observed Differences}
Most quantities do not differ at all between the original Belle
Software and the converted data using b2bii.

\begin{figure}
\centering
\vspace{-1em}
\begin{subfigure}[t]{0.45\textwidth}
  \hspace{-1em}
  \includegraphics[clip, trim=0cm 0.0cm 0cm 0cm,width=1.0\textwidth]{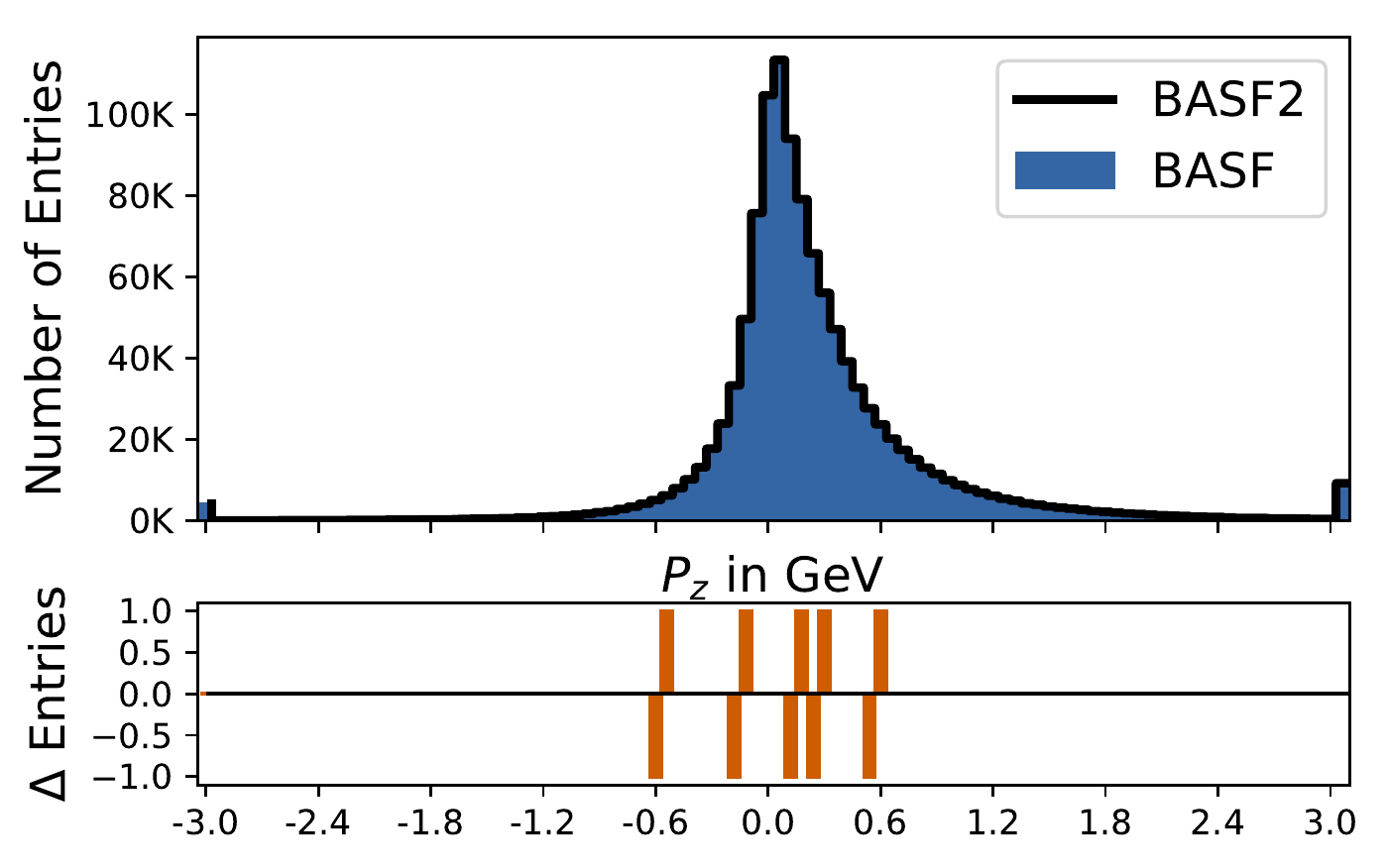}
  \vspace{-1em}
  \subcaption{Example of minor differences on recorded data: Momentum of Tracks in z direction exhibiting migration during the conversion due to numerical imprecision and special floating point values.}
  \label{fig:b2bii:Monitoring_MinorDifferences}
\end{subfigure}
\begin{subfigure}[t]{0.45\textwidth}
  \hspace{-1em}
  \includegraphics[clip, trim=0cm 0.0cm 0cm 0cm,width=1.0\textwidth]{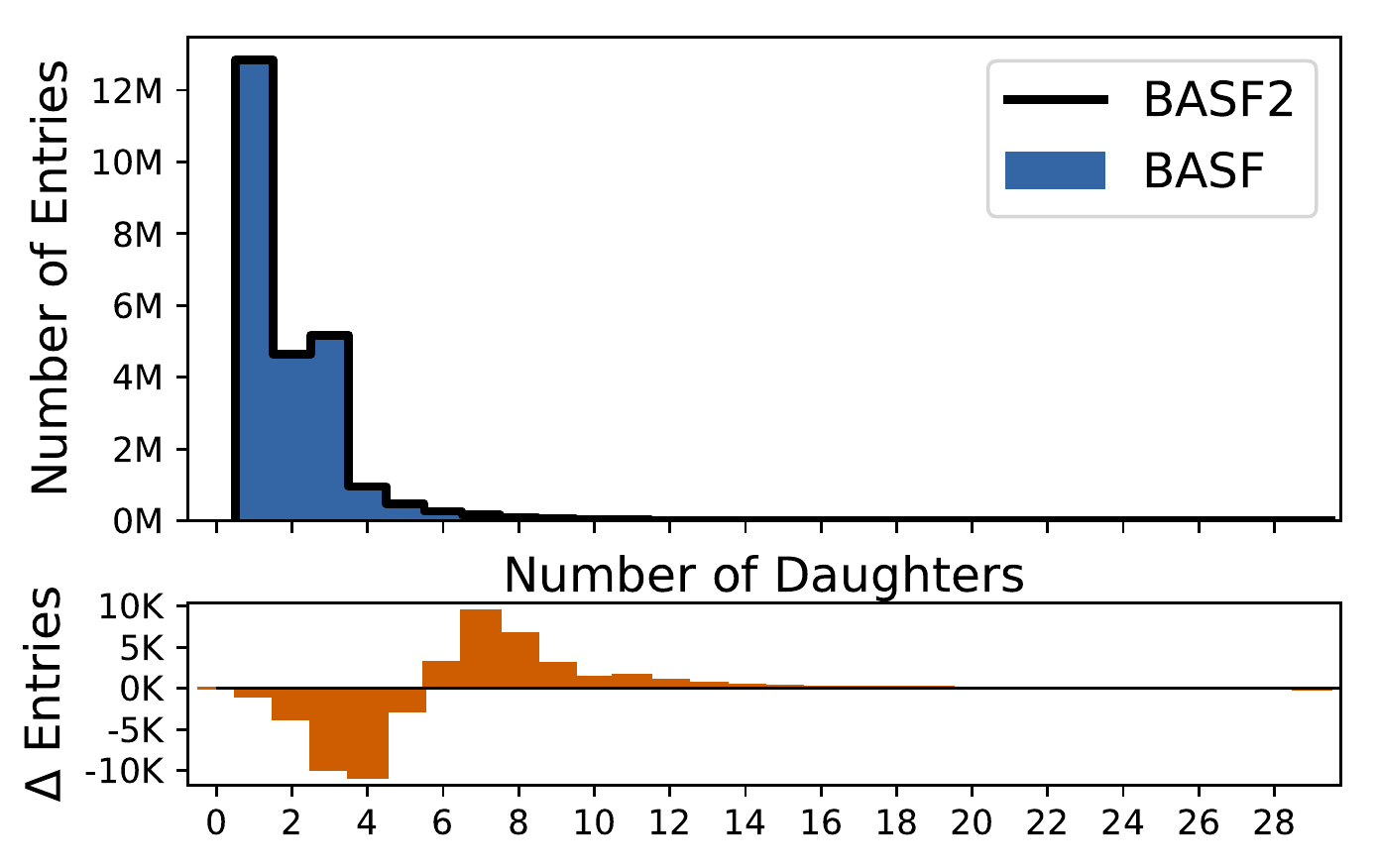}
  \vspace{-1em}
  \subcaption{Example of major differences on simulated events: The number of daughters of the Monte Carlo particle objects is shifted to smaller values because virtual photons are ignored during the conversion.}
  \label{fig:b2bii:Monitoring_MayorDifferences}
\end{subfigure}
  \caption{Comparison of \texttt{BASF} (Belle) and \texttt{b2bii} (Belle~II).  The leftmost (rightmost) bin represents the underflow (overflow) bin. The upper plots show the superimposed Belle (each component is shown individually) and Belle~II Monitoring Histograms (the total number of entries is shown as a black line). The lower plots show the differences between Belle and Belle~II, hence a positive (negative) difference means there are less (more) entries for the respective bin in the Belle~II Monitoring Histogram.}
\label{fig:b2bii:Monitoring}
\vspace{-1em}
\end{figure}

The observed differences between \texttt{BASF} and  \texttt{b2bii} were further investigated and either corrected or ensured to be harmless.
Minor differences occur due to small shifts caused by numerical imprecision leading to the migration of events between adjacent bins, especially for values near zero, and differences in the treatment of special floating point values such as infinity and NaN (Not a Number) leading to migration from the overflow/underflow bin to the bin including zero in rare cases (see \autoref{fig:b2bii:Monitoring_MinorDifferences}).

Further differences are found: in the PDG codes of the \texttt{MCParticle} object due to the recovery of the full 32~bit as mentioned above; the number of daughters of the \texttt{MCParticle} object due to the unconverted virtual photons occurring in nuclear interactions between the hadronic final state particles and the detector material (see \autoref{fig:b2bii:Monitoring_MayorDifferences}); and in all kinematic quantities of $V0$ and $\pi^0$ objects after the mass-constrained vertex fit caused by different software employed to fit the vertices.

\section{Conclusion}
\label{sec:b2bii:conclusion}
The Belle to Belle~II Conversion enables Belle~II physicists to analyze the dataset recorded by Belle using \texttt{BASF2}.
The conversion process was validated on a basic level by ensuring the same output for a large number of quantities. Differences which emerged were studied and explained.

In order to validate \texttt{BASF2} on a global level, physics analyses have been performed and compared to results published by the Belle collaboration \citep{master_felix, master_simon, master_fabian, master_judith, ThomasPhd, master_antonio}. Other measurements using the \texttt{b2bii} conversion are in preparation. 

Furthermore, \texttt{b2bii} is used to study the performance differences between the \texttt{Belle} and \texttt{Belle~II} experiment,
and to optimize the latter as soon as first data has been collected.

Finally, the conversion ensures the preservation of the legacy of the Belle experiment:
The full recorded dataset of approximately $\unit[1]{ab^{-1}}$ of data, which led to the verification of the CKM mechanism and the observation of tetra-quarks.

\bibliographystyle{unsrtnat}
\bibliography{short.bib}

\end{document}